# Multilayer Layer Graphene Nanoribbon Flash Memory: Analysis of Programming and Erasing Operation

Nahid M. Hossain[1], Md Belayat Hossain[2], Masud H Chowdhury[1]
[1]Computer Science and Electrical Engineering, University of Missouri – Kansas City, Kansas City, MO 64110, USA
[2]Applied Physics, Electronics & Communication Engineering, University of Dhaka, Bangladesh
Email: mnhtyd@mail.umkc.edu and masud@ieee.org

*Abstract-* Flash memory based on floating gate transistor is the most widely used memory technology in modern microelectronic applications. We recently proposed a new concept of multilayer graphene nanoribbon (MLGNR) and carbon nanotube (CNT) based floating gate transistor design for future nanoscale flash memory technology. In this paper, we analyze the tunneling current mechanism in the proposed graphene-CNT floating gate transistor. We anticipate that the proposed floating gate transistor would adopt Fowler-Nordheim (FN) tunneling during its programming and erase operations. In this paper, we have investigated the mechanism of tunneling current and the factors that would influence this current and the behavior of the proposed floating gate transistor. The analysis reveals that FN tunneling is a strong function of the high field induced by the control gate, and the thicknesses of the control oxide and the tunnel oxide.

## I. INTRODUCTION

Non-volatile flash memory is the most widely used memory devices in most of the portable and mobile electronics. Floating gate transistor (FGT) is the fundamental building block of the flash memory. Due to physical and material limitations further scaling of conventional silicon based floating gate transistors will not longer be possible in near future. Graphene and carbon nanotube (CNT) have emerged as highly potential future platforms for nonvolatile memories due to the extraordinary electrical, mechanical, thermal and physical properties of graphene and CNT. In our recent work [10], we presented the concept of a multilayer graphene nanoribbon (MLGNR) and Carbon nanotube (CNT) based floating gate transistor design for nonvolatile memory application. The conceptual design is shown in Figure 1 [10].

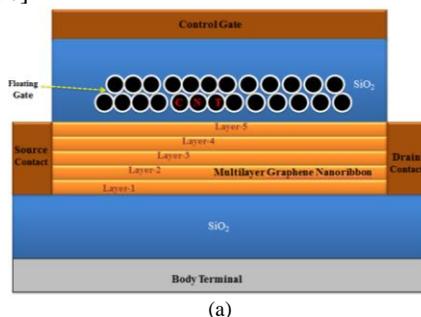

(a)
Figure 1: Conceptual layout of MLGNR-CNT based floating gate transistor

In the proposed floating transistor the logic '0' and '1' states are determined by the programming and erase operations respectively. Under the influence of a positive control gate voltage electrons are accumulated on the floating gate (programming) that translates to logic state '0'. A negative voltage applied at the control gate leads to the depletion of electrons (erase) that translates to the logic state '1'. The electron accumulation and depletion are accomplished by tunneling - a process by which an electron passes through a barrier without physical conduction path. Ideally, an insulating oxide barrier doesn't allow charge to pass through it. However, at high electric field and thin oxide thickness tunneling takes place. The tunneling effect becomes more prominent as device dimensions enter deep into nanometer scale while electric field strength is on the rise as supply voltage scaling is slowed. While for non-memory device tunneling through gate oxide is an undesired phenomenon the operation of floating gate transistors in nonvolatile memory is dependent on tunneling. Therefore, analyzing the tunneling mechanism of the proposed MLGNR-CNT based floating gate transistor is a critical part of our concept evaluation. In this paper, the tunneling mechanism during the programming and erase operations of the proposed transistor are investigated. The rest of the paper is organized as follows. Section II presents the fundamentals of the *Fowler Nordheim (FN) Tunneling,* which would be utilized to realize the operation of our proposed MLGNR-CNT floating gate transistor. Section III explains how the programming and erasing operations of the MLGNR-CNT floating gate transistor would depend on the tunneling mechanism. Section IV provides the results and analysis. Finally, Section V concludes the paper with a brief introduction to future work.

## II. FOWLER NORDHEIM (FN) TUNNELING

There are several mechanisms that allow charge to pass through insulating oxide. FN programming is achieved by applying a high voltage (around 15-20V for conventional CMOS FGT) at the control gate terminal while drain, source and bulk are grounded. For oxide layers thicker than 6nm, the tunneling current mechanism is explained by Fowler-Nordheim electron tunneling in MOS structures [6]. The Fowler Nordheim tunneling mechanism is widely used in non-volatile memory (NVM) for mainly three reasons: (i) tunneling is a pure electrical phenomenon, (ii) it requires very small programming current (< 1nA) per cell thus allowing many cells to be programmed at a time [11], and (iii) it allows very fast programming, which is a fundamental requirement for NVM technologies. FN tunneling is adopted in NAND flash memory, which is the most popular, dense and cost effective.

Channel hot electron (CHE) programming consists of applying a relatively high voltage (4~6 V for CMOS FGT) at the drain and a higher voltage (8~11 V for CMOS FGT) at the control gate while source and body are grounded. With this biasing condition a fairly large current (0.3 to 1 mA for CMOS FGT) flows in the cell and the hot electrons generated in the channel acquire sufficient energy to jump the gate oxide barrier and get trapped into the floating gate. Most NOR-type Flash memories utilize CHE programming. A third tunneling phenomenon, known as direct tunneling, can take place with ultra-thin oxide layers (2-5nm) at low or no biasing voltages [7]. There is a debate whether FN or direct tunneling is appropriate for 5nm~6nm oxide thickness because some researchers demand that FN tunneling is dominant for oxide thickness ≥ 4nm [1]. It is evident that FN tunneling is more dominant than direct tunneling when high electric field is applied. Most of the emerging NVM designs like the one presented in [12]-[15] propose to use a programming voltage around 15-20V. To be in coherence with the currently used programming voltage we anticipate using a programming voltage around 15V in our proposed design. That's why we mainly focus on FN tunneling based programing.

Fowler-Nordheim tunneling is the process where electrons tunnel through a barrier (gate oxide of a transistor) under the influence of a high electric field. It is a quantum mechanical process, where the electrons are injected by tunneling into the conduction band of the oxide through a triangular energy barrier (Figure 2). At high electric field band-bending takes place that results in apparent thinning of the barrier.

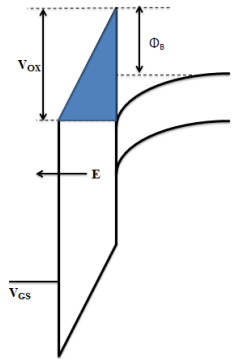

Figure 2: Fowler-Nordheim tunneling band diagram.

$$J_{FN} = \frac{k_1 E^2}{\phi_B} \exp(\frac{-k_2 \phi_B^{\frac{3}{2}}}{E}) \quad (1)$$

$$C_T = C_{FC} + C_{FS} + C_{FB} + C_{FD} \quad (2)$$

$$V_{FG} = (GCR \cdot V_{GS}) + \frac{Q_{FG}}{C_T} \quad (3)$$

The defining parameter for FN tunneling is the potential drop ($V_{OX}$) across control oxide. $V_{OX}$ should be greater than the barrier ($V_{OX} > \Phi_B$). The carriers (electrons) see a triangular barrier as in Figure 2. The tunneling current ($J_{FN}$) is dependent on the barrier ($\Phi_B$) seen by the carriers from the channel and the electric field across tunnel oxide as illustrated in (1). The dependence is dominated by the exponential term. From (1), it is observed that $J_{FN}$ depends exponentially on $\Phi_B$. Therefore, higher $\Phi_B$ leads to significantly lower $J_{FN}$. Higher electric field (E) leads to larger tunneling current.

### III. PROGRAMMING AND ERASING OF THE PROPOSED FLOATING GATE TRANSISTOR

Figure 3 shows the internal capacitive model of the proposed FGT. The dynamic behavior of the FGT is dependent on a critical parameter known gate control ratio (GCR). GCR is defined as the ratio of the capacitance $C_{FG}$ and the total capacitance ($C_T$) associated with the control and floating gate as shown in Figure 3. Here $C_T$ is given by (2). The voltage of the floating gate ($V_{FG}$) is dependent on the capacitances, control gate voltage ($V_{GS}$) and the accumulated charge on the floating gate ($Q_{FG}$) as shown in (3). During programming the source and the body terminals are connected to 0V. In silicon FGT, drain terminal is always connected to 0V during programming. In the proposed MLGNR-CNT FGT the drain is connected to a minimum voltage (50mV in this case) to increase the electron density in the graphene channel. In order to simplify the equations this very low drain voltage is considered to be 0V in the analysis. In conventional silicon FGT, the large substrate can supply enough charge to the floating gate. This is not the case in the proposed FGT. Besides very high control gate voltage ($V_{GS}$) is used in order to realize only the Fowler-Nordheim tunneling effects on the device operation while effects from other sources are minimized.

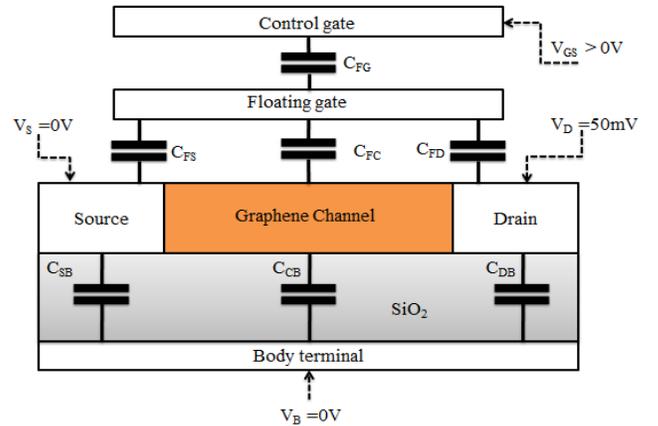

Figure 3: Terminal connections during programming.

During the programming let us consider that the initial charge $Q_{FG}=0$. With a voltage $V_{GS}=15V$ at the control gate of the MLGNR-CNT FGT and a GCR value of 0.6 the value of $V_{FG}$ would be 9V according to (3). This would lead to a voltage scenario as in Figure 4 that will result in a large tunneling current density ($J_{in}$) from the channel to the floating gate. On the other hand outward tunneling current density ($J_{out}$) is comparatively low because of the lower potential difference (15V-9V=6V) and thicker insulating oxide layer between the floating gate and the control gate. The thickness of the control oxide is always greater than the tunnel oxide. Therefore, $J_{in}$ is much higher than $J_{out}$ (Figure 4.). The relative strength of $J_{in}$ and $J_{out}$ are drawn along the Y-axis in Figure 4 for illustration.

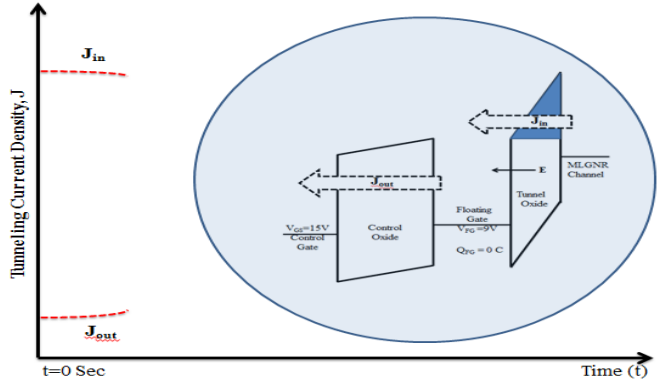

Figure 4: Tunneling current in time. Tunneling mechanism is shown in the insert at t=0 Sec.

As time progresses, more electrons are accumulated on the floating gate taking its potential below 9V (which was the potential of the floating gate for a given GCR and programing voltage when there was no charge accumulation). Negative charge accumulation on floating gate lowers $V_{FG}$, which leads to lower potential difference between the source and the floating gate. As a consequence $J_{in}$ decreases gradually as shown in Figure 5. However, during this process the potential difference between the floating gate and the control gate increases, which leads to higher $J_{out}$ as shown in Figure 5. As long $V_{FG}$ is larger than the potential difference between the control gate and the floating gate $J_{in}$ remains larger than $J_{out}$. At one time point $t = t_{sat}$ $J_{in}$ will be equal to $J_{out}$. The negative charge accumulated at $t_{sat}$ when $J_{in}=J_{out}$ represents the maximum charge that can be accumulated on the floating gate. This provides the range of programing voltage and time. The device will not useful as a nonvolatile memory cell for the range where $J_{in}<J_{out}$.

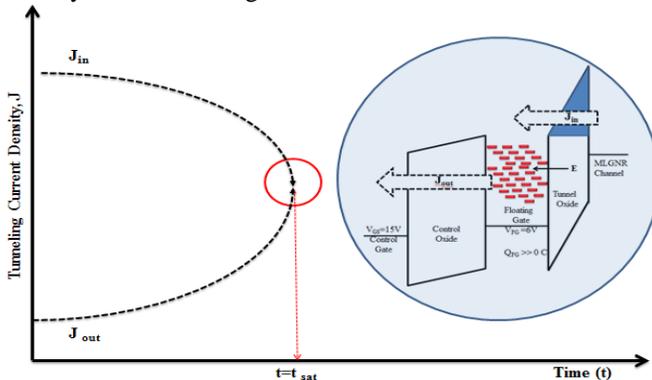

Figure 5: Tunneling current in time.

## IV. RESULT AND ANALYSIS

An important part of this analysis is to estimate $J_{in}$ and $J_{out}$, both of which can be modeled as FN tunneling current. One of the most widely approached FN tunneling current ($J_{FN}$) in the MOSFET structure is the Wentzel-Kramers-Brillouin (WKB) approximation as shown in (4) [3]. The parameters A and B depend on the work function or the barrier height ($\Phi_B$) at the interface between the tunneling oxide and the electron emitter and the effective mass of the tunneling electron $m_{ox}$. The work function is a property of the surface of the material. It depends on the crystal structure and the configurations of the atoms at the surface. A and B can be derived from FN plot ($J_{FN}/E^2$ vs. 1/E) as in [1]-[3]. Here, the induced electric field E is given by (5). By replacing E in (4) we get $J_{FN}$ as in (6). For source voltage $V_S$ =0V, $J_{FN}$ will be given by (7).

$$J_{FN} = AE^2 \exp\left[-\frac{B}{E}\right] \quad (4)$$

$$\text{Here, } A = \frac{q^3}{16\pi^2 h \Phi_B} \quad \text{and} \quad B = \frac{4}{3}\frac{(2m_{ox})^{\frac{1}{2}}}{qh}\Phi_B^{\frac{3}{2}}$$

$$E = \frac{V_{FG} - V_S}{X_{TO}} \quad (5)$$

$$J_{FN} = A\left(\frac{V_{FG} - V_S}{X_{TO}}\right)^2 \exp\left[-\frac{B}{\left(\frac{V_{FG} - V_S}{X_{TO}}\right)}\right] \quad (6)$$

$$J_{FN} = A\left(\frac{V_{FG}}{X_{TO}}\right)^2 \exp\left[-\frac{B}{\left(\frac{V_{FG}}{X_{TO}}\right)}\right] \quad (7)$$

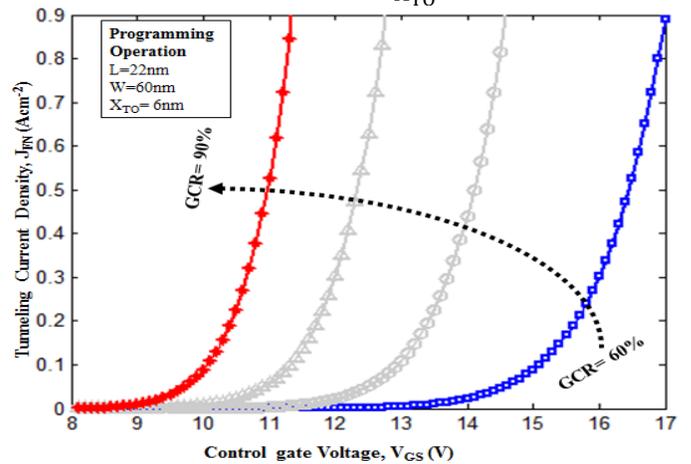

Figure 6: [Program] Fowler Nordheim (FN) tunneling current density ($J_{FN}$) versus Control gate voltage ($V_{GS}$) for four different GCR. $V_{GS}$ =8-17V.

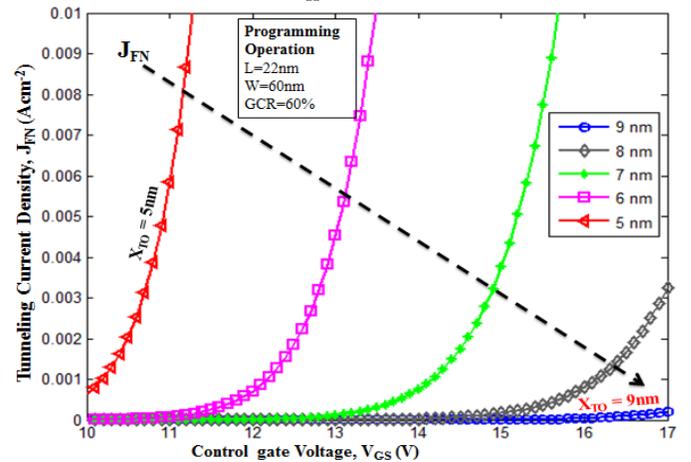

Figure 7: [Program] Fowler Nordheim (FN) tunneling current density ($J_{FN}$) versus Control gate voltage ($V_{GS}$) for five different tunnel oxide thickness ($X_{TO}$). GCR=60%, $V_{GS}$ =10-17V.

The subsequent paragraphs present the analysis of tunneling current during the programming and erasing operation of the proposed floating gate transistor based on the above models.

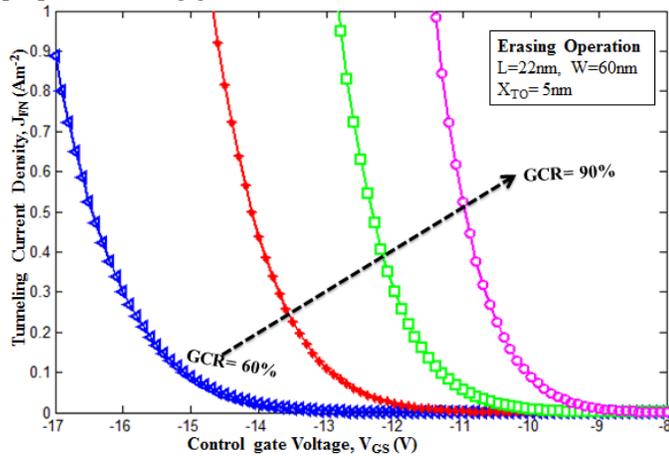

Figure 8: [Erasing] Fowler Nordheim (FN) tunneling current density ($J_{FN}$) versus Control gate voltage ($V_{GS}$) for four different GCR (%). $X_{TO}$=5, $V_{GS}$<0V.

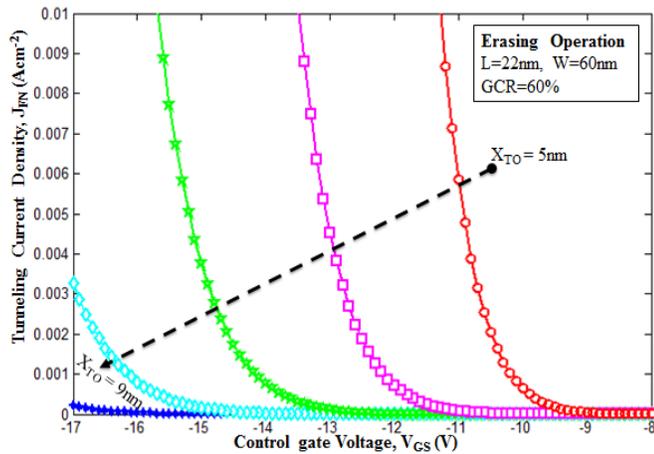

Figure 9: [Erase] Fowler Nordheim (FN) tunneling current density ($J_{FN}$) versus Control gate voltage ($V_{GS}$) for five different tunnel oxide thickness ($X_{TO}$). GCR=60%, $V_{GS}$<0V.

### a. Programming

Figure 6 shows the dependence of the FN tunneling current density ($J_{FN}$) on the control gate voltage ($V_{GS}$) for a given control gate coupling ratio (GCR). This set of graph is generated from equations (3) and (7). As can be seen $J_{FN}$ during programming increases with the increase of both the control gate voltage and GCR. Figure 7 shows the variation $J_{FN}$ with $V_{GS}$ for different tunnel oxide thickness ($X_{TO}$). It is observed that for a given $X_{TO}$, $J_{FN}$ increases with $V_{GS}$. However, $J_{FN}$ increases significantly when $X_{TO}$ is less than 7nm. According to ITRS 2011, semiconductor industry has already adopted 6nm tunneling oxide for 18-nm and 22-nm technology nodes. While 5nm tunnel oxide is predicted for 8-14nm technology nodes. Therefore, for technology nodes below 20nm, high tunneling current density will affect the reliability of the tunnel oxide.

### b. Erasing Operation

During the erasing operation a negative voltage would be applied at the control gate. We have performed the same set of analysis (as in Figure 6 and Figure 7) for the erasing operation. The same set of the Fowler Nordheim (FN) tunneling current density ($J_{FN}$) analysis is done for erasing operation. Figure 8 shows that $J_{FN}$ increases as the control gate voltage ($V_{GS}$) becomes more negative for a given GCR. Higher GCR leads to higher $J_{FN}$ because large control gate coupling will increase electron depletion rate from the floating gate to the MLGNR channel. Figure 9 shows the variation of $J_{FN}$ with $V_{GS}$ for different $X_{TO}$ during the erase operation. It is seen $J_{FN}$ increases with the increase of $V_{GS}$ in the negative direction for a given $X_{TO}$. The tunneling current increases significantly when $X_{TO}$ is less than 7nm similar to the programing operation.

## V. CONCLUSION AND FUTURE WORK

It is concluded for faster programming and erasing higher FN tunneling current density ($J_{FN}$) can be achieved by higher control gate voltage and scaling down the thicknesses of the control gate oxide and tunnel oxide. However, higher tunneling current will severely damage the oxide's reliability. Therefore, an optimization among these crucial parameters is recommended. Our future work will involve optimizing the supply voltage, tunneling current density and oxide thickness for optimum performance. Also, more accurate models for $J_{FN}$ and other electrical behaviors need to be developed.